\documentclass[aps,prd,showpacs,preprintnumbers,superscriptaddress]{revtex4}
\usepackage{amssymb}
\usepackage{bm}
\usepackage{enumerate}
\usepackage{amssymb}
\usepackage{amsmath}
\usepackage{feynmf}
\usepackage{slashed}
\usepackage{wrapfig}
\usepackage{graphicx}

\newcommand\be{\begin{equation}}
\newcommand\ee{\end{equation}}

\newcommand\bea{\begin{eqnarray}}
\newcommand\eea{\end{eqnarray}}

\newcommand\nn{ \nonumber \\}
\newcommand\oa{E_{A'}}
\newcommand\ma{m_{A'}}
\newcommand\ka{k_{A'}}
\newcommand\bka{{\bf k}_{A'}}
\newcommand\mn{m_N}

\newcommand\feynm{\mathcal{M}}

\begin{document}

\title{Constraints on Light Hidden Sector Gauge Bosons from Supernova Cooling}

\author{James B.\ Dent}
\affiliation{Department of Physics and School of Earth and Space Exploration, Arizona State University, Tempe, AZ 85287-1404, USA}

\author{Francesc Ferrer}
\affiliation{Physics Department and McDonnell Center for the Space Sciences, Washington University, St. Louis, MO 63100, USA}

\author{Lawrence M.\ Krauss}
\affiliation{Department of Physics and School of Earth and Space Exploration, Arizona State University, Tempe, AZ 85287-1404, USA}
\affiliation{Research School of Astronomy and Astrophysics, Australian National University, Canberra, AU, 2614}

\begin{abstract}
We derive new bounds on hidden sector gauge bosons which could produce new energy loss mechanisms in supernovae, enlarging the excluded region in mass-coupling space by a significant factor compared to earlier estimates.  Both considerations of trapping and possible decay of these particles need to be incorporated when determining such bounds, as does scattering on both neutrons and protons.  For masses and couplings near the region which saturates current bounds, a significant background of such gauge bosons may also be produced due to the cumulative effects of all supernovae over cosmic history.
\end{abstract}

\pacs{95.35.+d}


\maketitle

\section{Introduction}

Despite its phenomenological success, the Standard Model (SM) is known to be incomplete. Observations of neutrino masses and oscillations,
and several naturalness issues (stability of the electroweak scale, strong
CP problem) have motivated theoretical extensions, many of which involve the existence of new particles and forces.
Extensions of the Standard Model by inclusion of one 
or more Abelian groups are familiar in a number of contexts.  Additional 
$U(1)$ groups naturally arise from the breaking of GUT groups to the Standard 
Model, from attempts to resolve the fine-tuning of the Higgs mass, or from the possible existence of new
hidden sectors in nature (for a review see  \cite{Jaeckel:2010ni,Langacker:2008yv}).  
This latter alternative has elicited significant interest recently because it allows the possibility that new light gauge bosons associated with the new $U(1)$ groups may connect a
hidden sector to the observed sector, allowing for the existence of associated dark matter particles which may have enhanced couplings because of mixing between the new gauge bosons and photons.  

If the boson were massless 
then it would mediate long-range forces, which would then necessitate a very 
small coupling to SM matter or that the new group be sequestered from the SM. 
Massive gauge bosons are not as strongly constrained, but even for
additional $U(1)$ sectors that are not charged under the SM there will 
still arise a coupling to the SM through kinetic mixing.  That is, for a 
$U(1)'$ field strength $F'^{\mu\nu}$ with an associated gauge boson $A'$, 
and the hypercharge field strength $F^{Y\mu\nu}$, the Lagrangian density 
will contain the gauge invariant renormalizable terms~\cite{Holdom:1985ag}
\begin{eqnarray}
{\cal L} \supset \frac{1}{4}F'^{\mu\nu}F'_{\mu\nu} + \frac{\epsilon_Y}{2} 
F^{Y\mu\nu}F'_{\mu\nu}
\end{eqnarray}
If the $U(1)_Y$ is embedded in a GUT theory, Planck-suppressed operators
or loop-suppressed mixing from heavy split multiplets can induce
$\epsilon_Y \sim 10^{-8} - 10^{-2}$. In the context of string theory, the
possible range of $\epsilon_Y$ is much larger, with estimated values of
$\epsilon_Y\sim 10^{-17} - 10^{-5}$ from compactifications of the heterotic
string, or $\epsilon_Y\sim 10^{-12} - 10^{-3}$ in type II 
scenarios~\cite{Dienes:1996zr,Abel:2003ue,Abel:2008ai,Baumgart:2009tn}.
At low energies, the kinetic mixing can be removed by considering a shift in the photon field $A^\mu \rightarrow A^{\mu} + \epsilon A'^{\mu}$, inducing
$\epsilon$-suppressed electromagnetic interactions of the $A'$ with
strength $ e \epsilon$:
\begin{eqnarray}
{\cal L} \supset e \epsilon A'_{\mu}
J_{EM}^{\mu},
\label{eq:lagrangian}
\end{eqnarray}
where $\epsilon \equiv \epsilon_Y \cos \theta_W$, with weak mixing angle $\theta_W$.

As alluded to above, hidden sector models have garnered much attention recently because if there are dark matter particles in these sectors that are charged under a new dark gauge symmetry, the possibility exists for increased self interactions which could result, via mixing with the standard sector, in the possibility of an enhanced annihilation signature which was thought might explain anomalous cosmic ray data. Dark matter particles in the $U(1)'$ sector could
have a mass at the weak scale, with the $A'$ mass suppressed by 
$\sqrt{\epsilon}$ down to the MeV-GeV scale (there exists a rather larger literature on such enhancements, see for example \cite{Padmanabhan:2005es,Pospelov:2008jd,ArkaniHamed:2008qn,ArkaniHamed:2008qp,Hisano:2008ti,Hisano:2009rc,Baumgart:2009tn,Katz:2009qq,Dent:2009bv,Zavala:2009mi,Feng:2009hw,Buckley:2009in,Slatyer:2009yq,Feng:2010zp,Finkbeiner:2010sm,Hisano:2011dc,Slatyer:2011kg,Abazajian:2010zb,Zavala:2011tt,Abazajian:2011ak} for models and constraints). 

Of course, if the new gauge bosons are sufficiently light, they can also produce observable signatures, using direct and indirect terrestrial probes.  Beam dumps, for example, which are sensitive to the possibility of new light penetrating particles such as axions provide stringent constraints \cite{Bjorken:2009mm, Krauss:1986bw}.  In addition, one must consider the impact of new light gauge bosons on sensitive atomic probes such as the anomalous magnetic moment of the muon \cite{Pospelov:2008zw}.

Supplementing these constraints on a hidden sector are constraints from astrophysics. If hidden sector gauge particles mix with photons and then escape from a star, then this new energy loss mechanism can dramatically
affect not only stellar structure but also stellar evolution.  Perhaps nowhere is this more dramatic than in the case of core collapse supernovae. 
 As a result, supernova cooling constraints
(namely  the energy loss observed from SN1987a) have long been employed in 
order to constrain particle couplings and masses 
\cite{Iwamoto:1984ir,Turner:1987by,Brinkmann:1988vi,Ellis:1988aa}.  

Early work was concerned principally with constraints on nearly massless axions emitted in nucleon-nucleon bremsstrahlung 
processes inside of  the hot ($T\sim 30 MeV$) neutron star born from an 
associated supernova.  One can adapt this line of inquiry to constrain 
the masses and mixings of a hidden sector $U(1)'$ gauge boson, $A'$, 
which may couple to Standard Model-sector charged particles via the previously mentioned kinetic 
mixing.  An estimate 
of the energy loss of such a process was given in \cite{Bjorken:2009mm}, 
but, as is mentioned in that work, this estimate was expected to be accurate at best at the order of magnitude level.  

It is the purpose of this work to provide a more thorough accounting of this process and thus derive more accurate constraints on the masses of such 
particles, $m_{A'}$, and their couplings, $\epsilon$.  
Because models of interest for dark matter involve gauge boson masses which exceed $1$ MeV, we restrict ourselves to this range here.  Moreover, for much lighter masses, other considerations arise.  In particular, matter effects as particles traverse the high density collapsing core can be large enough to produce oscillation phenomena reminiscent of that for neutrinos which can cause enhanced trapping of such particles and thus may obviate the constraints we derive on heavier gauge bosons~\cite{Redondo:2008aa}.  Dark
matter particles in the $U(1)'$ sector are generically heavy and would not contribute to the cooling of 
the supernova. In scenarios where dark matter particles have masses below
$\sim 10$ MeV, that could explain the 511 keV line emission from the
galactic bulge~\cite{Pospelov:2007mp,Pospelov:2008jd,Pospelov:2008zw}, 
additional cooling channels could be 
present~\cite{Mohapatra:1990vq,Fayet:2006sa}, which we do not consider below.

\section{Free Streaming}

We are interested in exclusion regions in the $\epsilon-m_{A'}$ plane.  
What one expects is that there will be an upper bound on $\epsilon$ 
above which the vector particle production will exceed the bounds from 
SN1987a, extending up to a value when the vector particles would be strongly enough coupled to be trapped inside the supernova core.  This is what we will call the free streaming region.  In the first 
sub-section we will outline calculations for the process 
$p + p \rightarrow p + p + A'$, and then extend this to include the $p + n 
\rightarrow p + n + A'$ process as well.  

\subsection{The proton only case}

The central quantity to calculate is the emission rate of the vector particles, 
which is labeled $Q_{ij}$ where the subscripts stand for the possible 
nucleons participating in the emission process.  
In the hot medium with temperature of a few MeV and densities typical of
the accretion disk, the nucleons are nondegenerate and nonrelativistic. The
main emission process is the nucleon-nucleon-dark boson bremsstrahlung.
In studies of axion emission one usually neglects the mass of the
emitted particle. Since our dark gauge boson can have a mass comparable
to the temperature of the SN medium, this approximation is not justified
and we follow the kinematical analysis of~\cite{Giannotti:2005tn}.
The details of this 
somewhat lengthy calculation in the one-pion-exchange (OPE) approximation
are given in the appendices.  If we 
take the expression for $Q_{pp}$ given in Eq.(\ref{qpp}), and use the fact that the $A'$ couples to protons and charged pions with strength 
$g_{\pi} = g_p \equiv e \epsilon$, we can write:
\be
Q_{pp} = 
\frac{\alpha_\pi^2 T^{2.5} \rho^2}{32 \pi^{1.5} m_N^{5.5}} 2 f_{pp}^4 e^2 
\epsilon^2 {\cal I}_k (y, q),
\label{eq:q}
\ee
where ${\cal I}_k$ denotes the phase space integration 
in terms of the dimensionless variables $y=m_\pi^2/m_NT$, and $q=m_{A'}/T$.
We have also replaced the baryon number density, $n_B$, by
its mass density through $n_B = \rho/m_N$.
For the pion-nucleon coupling we take $f_{pp} \approx 1$, resulting in 
$\alpha_\pi \approx 15$ in Eq.~(\ref{eq:q}). 
In our calculations we will use the typical 
values $\rho = 3 \times 10^{14} {\rm g/
cm^3} \equiv 3 \times \rho_{14}$ and 
$T \sim 30$ MeV
for the density and temperature in the interior of a supernova. 
We then constrain the parameter space by imposing that
the luminosity due to vector emission must be less than:
\be
{\cal L}_v \leq 10^{53} {\rm erg/s} \approx 4.1 \times 10^{37} {\rm MeV}^2,
\label{eq:lv}
\ee
which is roughly the energy lost in neutrinos~\cite{Raffelt:1996wa}.

Since our expression for $Q_{pp}$, Eq.~(\ref{eq:q}), 
is the luminosity per unit
volume, we need to integrate it over the volume where the vectors are 
emitted.  We then assume that $T\sim 30$ MeV holds within a central spherical
region of 1 km in radius (if we take the SN to have a mass $\sim 1.5 M_\odot$, 
then it would have a total radius of $\sim 13$ km assuming the same constant 
density).



Putting all this together, we find that the luminosity due to vector emission
\be
Q_{pp} V = 1.05 \times 10^{48} \rho_{14}^2 T_{MeV}^{2.5} r_{km}^3 
\epsilon^2 {\cal I}_k (y, q) \; {\rm 
MeV}^2 \approx 4.7 \times 10^{52} \epsilon^2 {\cal I}_k (y, q) \; {\rm 
MeV}^2,
\ee
cannot be larger than ${\cal L}_v$ in Eq.~(\ref{eq:lv}).

So our constraint on the coupling can be written as:
\begin{equation}
\epsilon  \leq \sqrt{\frac{3.9 \times10^{-11}}{\rho_{14}^2
T_{MeV}^{2.5} r_{km}^3 {\cal I}_k(y,q)}} 
\approx \sqrt{\frac{8.8\times10^{-16}}{{\cal I}_k (y,q)}}, 
\label{eq:constraint}
\end{equation}
and we then generate an exclusion bound with $y$ fixed, while varying $q$ for
different masses $m_{A'}$.  This will give an upper bound on the coupling $\epsilon$ in terms of $m_{A'}$.

\subsection{Including neutron processes}

Next we will include processes with both protons and neutrons.  
The expression for emission due to the process 
$p + n \rightarrow p + n + A'$ is given in Eq.~(\ref{qpn}).  
This situation is slightly more complicated as we now have five integrals contributing in the 
phase space integration.  
We can use Eq.~(\ref{eq:constraint})
to find the limits, if we add a multiplicative factor of 8 in $Q$. This 
comes about as follows: there is a factor of $4 = (\sqrt{2})^{4}$ from isospin
requirements that the neutron-proton coupling to a charged pion is $f_{pn}=\sqrt{2} f_{pp}$;
there is no longer a symmetry factor of $1/4$, since proton and neutron are not 
indistinguishable; finally, we had a factor of 2 from $g_\alpha^2 + g_\beta^2$
in front of ${\cal I}_k$, but now $g_\beta=0$. The results are shown
in Fig.~\ref{fig:totaldecay}.

\section{Including decay}

\begin{figure}
\begin{center}
\includegraphics{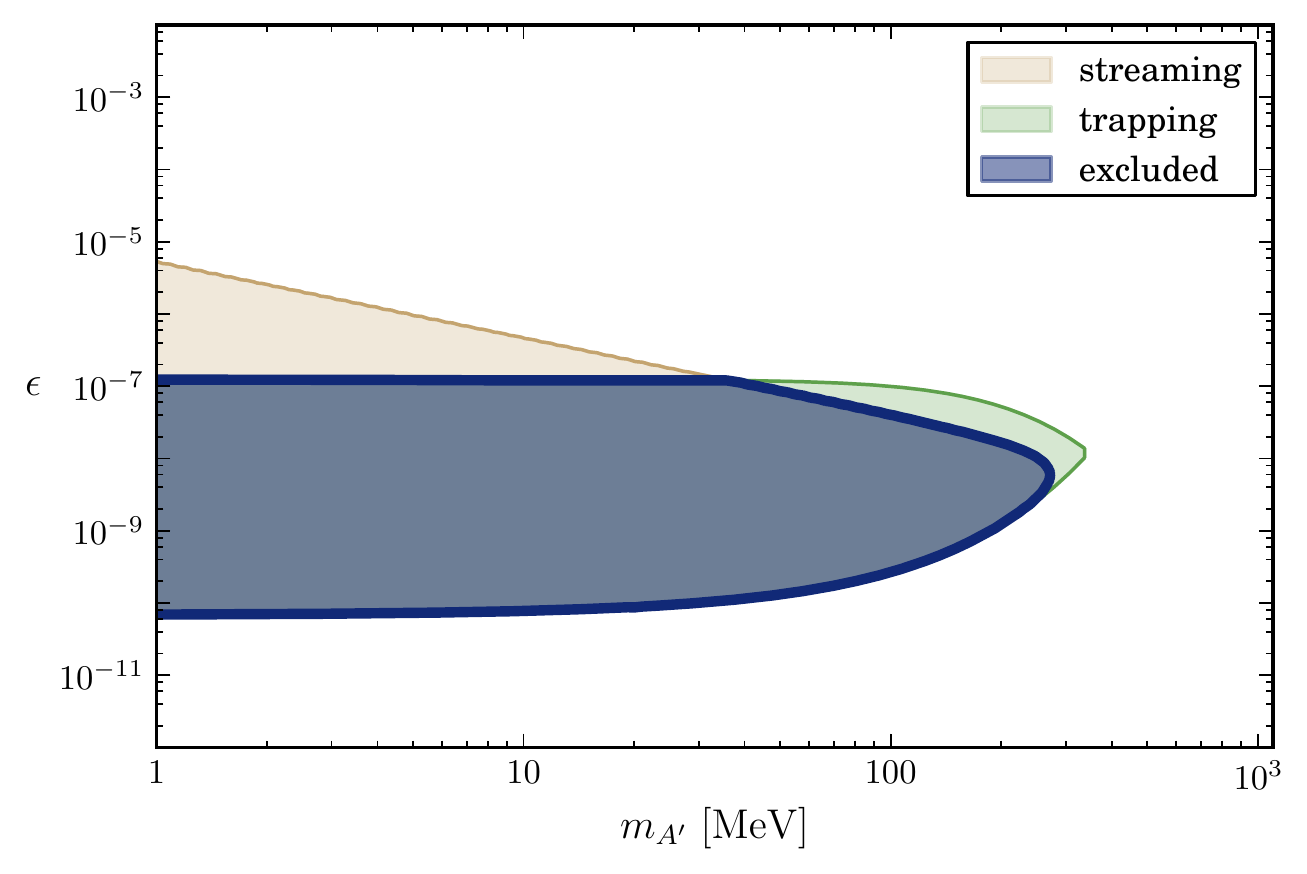}
\end{center}
\caption{The total excluded region, adding contributions from both the $pp$ and $pn$ processes}.
\label{fig:totaldecay}
\end{figure}

Naively, larger couplings that do not satisfy Eq.~(\ref{eq:constraint}) 
would be excluded.
But, the free streaming limit is not the only region in parameter space where 
constraints will arise. In general, due to the couplings in 
Eq.~(\ref{eq:lagrangian}),
the dark gauge boson decays back into
leptons and other SM particles within a distance:
\be
l_0 = \frac{3 E_{A'}}{N_{eff} m_{A'}^2 \alpha \epsilon^2}.
\label{eq:l0}
\ee
In order to implement this, 
we will follow the strategy outlined in \cite{Bjorken:2009mm} 
and modulate the emission amplitude with an 
exponential damping factor,
\be
{\rm{e}}^{-10\;km/l_0}.
\ee
The assumption is that the products of the decay remain within the SN core
and do not contribute to the cooling~\footnote{This assumption might
break in scenarios where $A'$ can decay into hidden sector matter with
a large branching ratio.}. As the mixing parameter $\epsilon$
increases, the decay happens earlier and the gauge bosons are not effective
at cooling, as seen in Fig.~\ref{fig:totaldecay}.  
Therefore we would expect that 
above some value of $\epsilon$, the emission constraints will no longer hold,
and we will find a lower bound on the coupling, creating a region bounded above and 
below once decay is included with free streaming.

In terms of $r_{km}$ and the decay length, Eq.~(\ref{eq:l0}), 
the exponential parameter can be written as:
\be
{\rm{e}}^{-2.4 \times 10^{14} \frac{r_{km} N_{eff} q^2 \epsilon^2}{x y}},
\ee
where $x=E_{A'}/T$.

This exponential factor enters the phase space integration, 
and as a result, the
function ${\cal I}(y,q)$ becomes a function ${\cal I}^{dec} (y, q,
\epsilon, r_{km})$, but we can still use 
Eq.~(\ref{eq:constraint}) with ${\cal I}^{dec}$.  
When generating the plots, we took $N_{eff}=1$, which 
corresponds to the situation where only the decays $A'\rightarrow e^+ e^-$
are possible. We also take $r_{km} = 10$ for the typical distance where
decay products are trapped, and, as in the previous section, we assume that
vector particles are produced within the inner km.


\section{Trapping limit}

Thus far we have examined the energy loss due to vectors which are emitted and subsequently decay outside of the supernova.  However, as is well known, above a certain coupling, these particles can be trapped within the supernova.  This will create a lower bound on the coupling $\epsilon$, 
above which the trapping will not allow vector emission to contribute to cooling the supernova.



If the new particles generated in the nucleon interaction processes 
subsequently interact strongly enough they will thermalize, and
will be emitted from a spherical shell where the optical depth is
roughly $\tau \approx 2/3$. If the temperature is $T_x$, the luminosity
will be given by the Steffan-Boltzmann law:
\be
L = 4 \pi r^2 \sigma T_x^4.
\label{eq:steffanboltzmann}
\ee
Here $\sigma$ is the Steffan-Boltzmann constant, which is $\pi^2/60$ for
photons and 
\be
\sigma \equiv \frac{g \pi^2}{120},
\ee
for a new particle with $g$ degrees of freedom.

In principle, one should determine $r$ and $T_x$, but 
following \cite{Raffelt:1996wa} we assume
$r \approx 10$ km. This is a good approximation, 
because the density of the protoneutron 
star falls abruptly near that radius.  

The requirement that the luminosity in new particle emission be less than
\be
L \lesssim 10^{53} {\rm erg/s} \approx 4.1\times10^{37} {\rm MeV^2}
\ee
translates into a bounded relation between the temperature and the coupling
\be
T_x \lesssim 11.4\; g^{-1/4}\: {\rm MeV}.
\label{eq:trappingconst}
\ee


To calculate $T_x$ one needs a model for the temperature
and density above the settled inner SN core, for which we turn to 
\cite{Turner:1987by}:
\begin{align}
\rho(r)&=\rho_R \left(\frac{R}{r}\right)^n \nn
T(r) &= T_R \left(\frac{\rho(r)}{\rho_R}\right)^{1/3},
\label{eq:snmodel}
\end{align} 
where $\rho_R = 10^{14} {\rm g/cm^3}$, $R=10$ km, $T_R = 10-20$ MeV, and
$n \sim 5$ is a relatively large number.

As shown below, given a model, 
the opacity $\kappa (\rho, T)$ can be expressed in terms
of $r$. It is then straightforward to compute the optical depth,
\be
\tau\left(r_x\right) = \int_{r_x}^\infty {\kappa \rho dr}.
\label{eq:odepth}
\ee

One then solves for $r_x$ from $\tau \left(r_x\right) \approx 2/3$, and uses
Eq.~(\ref{eq:snmodel}) to find $T_x$. The temperature just found is used
in Eq.~(\ref{eq:steffanboltzmann}) or compared to 
Eq.~(\ref{eq:trappingconst}).


To find the
 opacity as a function
of $\rho$ and $T$, one starts from the reduced Rosseland mean opacity:  
\begin{align}
\frac{1}{\kappa_x \rho} & \equiv \frac{1}{4 a T^3} \int_{m_x}^\infty {d
\oa l_{\oa} \beta_{\oa} \partial_T B_{\oa} } \nn
&=\frac{15 g}{8 \pi^4 T^5} \int_{m_x}^\infty {d
\oa l_{\oa} \frac{\oa^2 \left(\oa^2 - m_{A'}^2 \right) {\rm e}^{\oa
/T}}{\left({\rm e}^{\oa/T}-1\right)^2}},
\end{align}
where we have used $a = \pi^2/15$, the velocity $\beta_{\oa} =\sqrt{1
-(\ma/\oa)^2}$, and the definition of $B_{\oa}$ is given by
\begin{eqnarray}
B_{\oa} = \frac{g}{2\pi^2}\frac{\oa^2(\oa^2-m^2)^{1/2}}{e^{\oa/T}-1}.
\end{eqnarray}

Since the production of bosons involves a Bose stimulation factor, we need
to add a factor of $\left(1-{\rm e}^{-\oa/T}\right)$ under the integral
above. This gives the {\it reduced} opacity (often denoted as $\kappa^*$), which is the
one that we have been using. Dropping the superscript ${}^*$, we find that:
\be
\frac{1}{\kappa_x \rho} = \frac{15 g}{8 \pi^4} \int_q^\infty {dx
l_x \frac{x^2 \left(x^2 - q^2 \right) {\rm e}^{2 x}}{\left({\rm e}^x -1
\right)^3}},
\label{eq:opacityv}
\ee
which reduces to the expression for axions~\cite{Raffelt:1996wa} 
for $g=1$, and $q=m_x/T=0$.

In the corresponding calculation for a massless axion, the mean
free path can be obtained as follows: starting from the energy-loss 
rate, $Q_x$, one removes the phase-space integral
$\int_0^\infty{d\oa 4 \pi \oa^2/(2 \pi)^3}$, and a factor of $\oa$, since
$Q_x$ is an energy loss rate.
Additionally a factor of ${\rm e}^{\oa/T}$ must be included 
to account for the detailed-balance relationship, 
and this gives $l_{\oa}^{-1}$. 

In our case, we have to take into account that the boson under consideration is
massive.  Therefore, the vector boson phase-space integral,
is $\int{d^3 {\bm p}}=\int{4 \pi p^2 dp}$, with 
$p=\sqrt{\oa^2 -\ma^2}$, and $ 4 \pi p^2 dp = 4 \pi \sqrt{\oa^2 -\ma^2} 
\oa d\oa$.

Hence, the inverse mean free path is obtained by removing
from the luminosity, $Q_{ij}$, the factor:
\be
\int_{m_x}^\infty{d \oa \frac{4 \pi \oa \sqrt{\oa^2 -\ma^2}}{(2 
\pi)^3}} \oa = T^4 \int_q^\infty{dx \frac{4 \pi x^2 \sqrt{x^2 - q^2}}{( 2 \pi)^3}},
\ee
and adding the detailed balance term.

Our expression for $Q_{ij}$ is once again given by Eq.~(\ref{qpp}) for the
$p+p\rightarrow p + p + A'$ case, and in Eq.~(\ref{qpn}) for the 
$p+n\rightarrow p + n + A'$ bremsstrahlung.  We can write them both as follows:
\begin{align}
Q_{ij} &= \frac{\alpha_\pi^2  \eta_{ij}
T^{2.5} \rho^2}{32 \pi^{1.5} m_N^{5.5}} f_{ij}^4 e^2 
\epsilon^2 \nn
&\times \int{du dv dx \sqrt{uv} {\rm e}^{-u} \frac{\sqrt{x^2-q^2}}{x} 
\delta(u-v-x) I_{ij}},
\end{align}
where $I_{ij}$ are functions defined in the appendix.
The factor $\eta_{ij}$ is equal to $2$
for $pp$ scattering, and $\eta_{pn}=16$, with the usual 
isospin and non-identical particles enhancement
in $pn$ scattering. 

Next, we perform the $u$-integration with the help of the delta function, 
and we remove a factor of $\oa= T x$ from the
$x$-phase-space integration. Additionally we must include the detailed
balance term, which cancels an ${\rm e}^{-x}$ that appeared after the delta
function was used.

Once all of these machinations are completed, the mean free path 
corresponding to each channel is found to be
\begin{align}
l_{\oa}^{-1}&=\frac{\alpha_\pi^2  \eta_{ij}
T^{2.5} \rho^2}{32 \pi^{1.5} m_N^{5.5}} f_{ij}^4 e^2 
\epsilon^2 T^{-4} \frac{(2 \pi)^3}{4 \pi} \nn
&\times \frac{1}{x^3} \underbrace{\int_0^\infty {dv \sqrt{(v+x)v} {\rm e}^{-v} 
I_{ij}}}_{\equiv {\cal S}_{ij}(x ; q, y)}.
\end{align}

We can now plug this expression into Eq.~(\ref{eq:opacityv}), and manipulate
it to find
\be
\kappa_{ij} = \frac{\alpha_\pi^2 \sqrt{\pi} f_{ij}^4 e^2 
\epsilon^2}{16 m_N^{5.5}}
\hat{\kappa}_{ij} \frac{\rho}{T^{1.5}},
\label{eq:kappav}
\ee
where we defined the dimensionless opacity:
\be
\hat{\kappa}_{ij}^{-1} \equiv \frac{15 g}{8 \pi^4} \frac{1}{\eta_{ij}}
\int_q^\infty{dx \frac{x^2 \left(x^2-q^2\right) {\rm e}^{2x}}{\left({\rm e}^x
-1 \right)^3} \frac{x^3}{{\cal S}_{ij} (x)}},
\ee
and ${\cal S}$ is defined above. We find the total contribution from
both channels 
using the fact that inverse opacities add: $\kappa^{-1} = 
\kappa_{pp}^{-1}+\kappa_{pn}^{-1}$.

One then defines the quantity $\tau_R \equiv \kappa_R \rho_R R$. 
Substituting $\rho_R$ and $T_R$ in
Eq.~(\ref{eq:kappav}), we can find $\tau_R$, and 
\be
\kappa \rho R = \tau_R \left(\frac{\rho}{\rho_R}\right)^2 
\left(\frac{T_R}{T}\right)^{3/2}.
\ee 

Finally the opacity is found from Eq.~(\ref{eq:odepth}) to be
\be
\tau_x \left(r_x\right) = \frac{\tau_R}{\frac{3}{2} n -1}
\left(\frac{T_x}{T_R}\right)^{9/2 -3/n}.
\ee

This expression can now be used to bound the coupling $\epsilon$ as a function of the dark gauge boson mass $m_{A'}$ by requiring $\tau_x \lesssim 2/3$.  For the case considered in this section, of trapping without including decay, we find numerically the exclusion region shown in Fig.~\ref{fig:totaldecay}.  

\begin{figure}[h!]
\begin{center}
\includegraphics{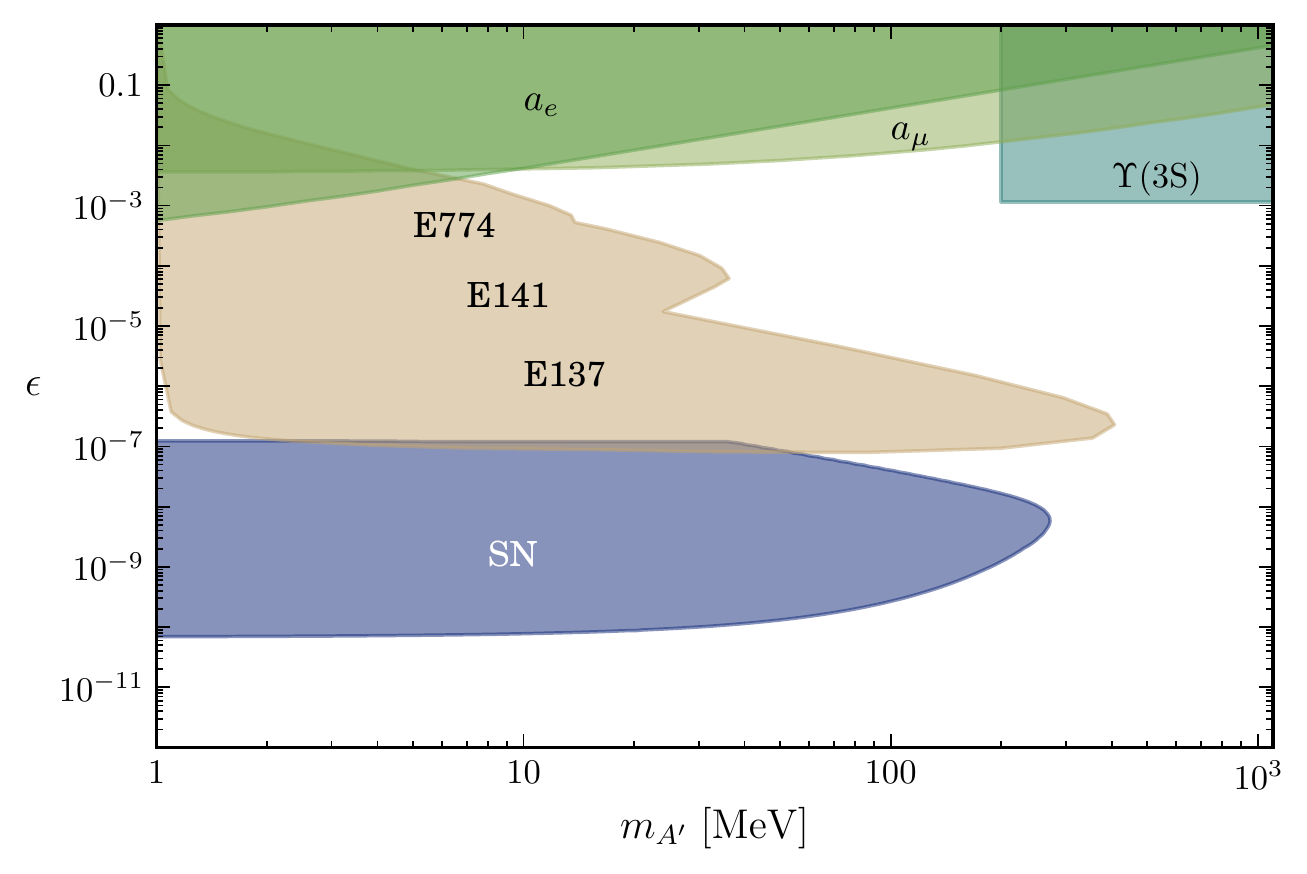}
\end{center}
\caption{The total excluded region for MeV to GeV dark gauge boson masses is shown including the supernova bounds calculated in this work, shown in blue, along with constraints due to other sources such as beam dump experiments E774, E141, E137, contributions to the anomalous magnetic moment of the muon and the electron, $a_{\mu}$ and $a_{e}$, and BABAR bounds from
upsilon $\Upsilon(3S)$ decays.}
\label{fig:MeVTeV}
\end{figure}

In a more realistic scenario one must take into account both trapping as well as decay.  This will lead to an exclusion region which is the intersection of the regions from trapping and decay alone, as either process will ensure that no additional energy loss will occur.  For small masses and stronger couplings, trapping constrains the excluded region of parameter space.  Conversely, for larger masses and smaller couplings, decay becomes most important.  The combined region is shown in Fig.~\ref{fig:totaldecay}.  

In Fig.~\ref{fig:MeVTeV} we display \footnote{We thank J. Redondo for providing us with the non-supernova constraints shown in Fig.~\ref{fig:MeVTeV}.} the combined region along with constraints due to other sources such as beam dump experiments E774, E141, E137, contributions to the anomalous magnetic moment of the muon and the electron, $a_{\mu}$ and $a_{e}$, and BABAR bounds from upsilon 
$\Upsilon(3S)$ decays. 

\section{A Diffuse Dark Gauge Boson Background?}

With an average event rate of approximately $10^{-2}$ yr$^{-1}$ per galaxy, there have been over $ 10^{18} $ supernova explosions over cosmic history within our observable horizon.  As was recognized even before SN1987a allowed more accurate estimate of neutrino production rates in supernova, and before recent supernova surveys allowed a more careful determination of event rates, the cumulative effect of such supernovae can produce a significant diffuse background of particles \cite{Krauss:1983zn}.   In particular a neutrino background in the MeV range with a flux in the range of $ \approx 10^2$cm$^{-2}$sec$^{-1}$ has been predicted.   If a new dark gauge boson exists which is not ruled out by the arguments presented here, but which nevertheless contributes significantly to supernova cooling, a similar background of such particles would be expected.  Indeed, similar considerations have recently been recognized for a possible diffuse axion background from supernovae \cite{Raffelt:2011ft}. Because these particles are massive, the dominant contribution to energy loss will come from their mass as long as it exceeds the mean temperature near the surface of the supernova neutrinosphere.  As a result we would expect a diffuse background today of dark gauge particles of mass $m_{A'}$ as large as  
$\approx 10^2$ (MeV$/ m_{A'}$) cm$^{-2}$sec$^{-1}$.   Their net contribution to the overall mass density of the universe would be negligible, but depending upon their interactions, such a background might be detectable.  Moreover, if these particles are not stable, but decay into Standard Model particles, they would contribute to a diffuse cosmic ray background.  We are therefore currently examining whether various direct and indirect constraints may be relevant for the possible detection of such a background \cite{inprogress}.

\section{Discussion}

The calculations described here produce more accurate, and more importantly, significantly enhanced constraints on light gauge bosons that may arise from hidden sector models which predict novel types of dark matter.   Mixing with photons produces results in bremsstrahlung production by scattering of both protons and neutrons, and a careful consideration of the resulting constraints from observations of SN1987a, involving both considerations of trapping and possible decay allow us to rule out a significantly larger region of masses and mixings than was previously estimated \cite{Bjorken:2009mm}, extending a constraint on an effective coupling of less than $10^{-10}$ for a significant range of gauge boson masses.   We also note that for masses and mixings near those that saturate the supernova cooling bounds, a significant diffuse background of dark gauge bosons will be produced by the cumulative effects of all supernovae over cosmic history.  Whether such a background might be detectable is of some interest, and is the subject of ongoing investigations.

\section{Acknowledgements}

We would like to thank H. Mathur, M. Ogilvie, J. Redondo, and T. Vachaspati 
for helpful discussions.   LMK and JBD acknowledge support from the 
Department of Energy and Arizona State University.  FF was supported in
part by the U.S. DOE under Contract No. DE-FG02-91ER40628 and the NSF
under Grant No. PHY-0855580.

\appendix
\numberwithin{equation}{section}
\section{Amplitude for the process $ p + p \rightarrow p + p + A'$}

Nucleon-nucleon bremsstrahlung is the primary mechanism for dark gauge
boson emission from the inner core of the SN. Like in the corresponding
axion calculation, we calculate the rate for this process in the
one-pion-exchange (OPE) approximation, which is sufficiently reliable
for our present purpose~\cite{Turner:1989wa} (although it could overestimate
the emission rates by a small factor 
at lower temperatures~\cite{Hanhart:2000ae}).
Since the hidden sector boson only couples at tree level to SM particles with electric charge, we need only 
consider nucleon scattering processes which involve protons.  
We have 
eight diagrams for the OPE proton scattering processes 
with vector bremsstrahlung, 
$p+ p\rightarrow p + p + A'$ and five diagrams for the $p + n\rightarrow p + n + A'$ process, where 
$A'$ is the hidden sector $U(1)'$ gauge boson with mass $m_{A'}$, 
and we will take all nucleon masses to be $m_N$.  In this Appendix and in Appendix B we will examine the process $p+ p\rightarrow p + p + A'$, with a treatment of the process $p+ n\rightarrow p + n + A'$ given beginning in Appendix C.

\begin{figure}
\begin{center}
\includegraphics[width=0.7\textwidth]{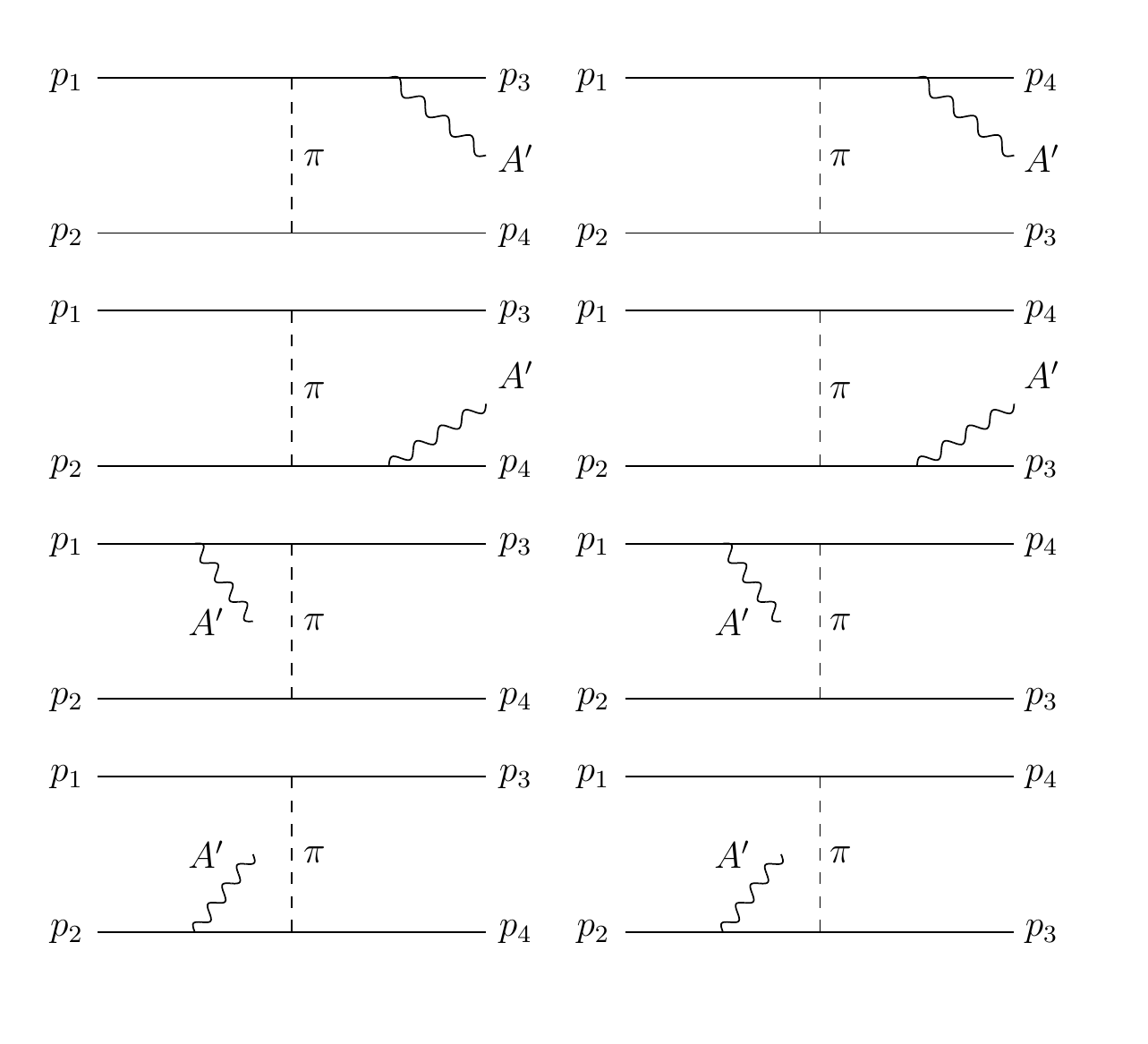}
\end{center}
\caption{Diagrams for the process $p + p \rightarrow p + p + A'$.  Following \cite{Brinkmann:1988vi}, these are labelled from top left to bottom left as $a, b, c, d$, and from top right to bottom right as $a', b', c', d'$.}
\label{fig:feynman}
\end{figure}

Following \cite {Brinkmann:1988vi} these diagrams, as shown in Fig.(\ref{fig:feynman}), are designated $a, b, c, d, a', b', c', d'$, which gives a total of sixty-four terms from the squared matrix element
\begin{eqnarray}
\sum_s \mathcal{M}\mathcal{M}^{\dag} = \sum_s |\mathcal{M}_a + \mathcal{M}_b + \mathcal{M}_c + \mathcal{M}_d -\mathcal{M}_a' -\mathcal{M}_b' -\mathcal{M}_c' -\mathcal{M}_d'|^2 
\end{eqnarray}
The plus sign is for the direct $t-$channel diagrams and the minus sign is for the exchange $u$-channel diagrams.  
As required by gauge invariance, the amplitude satisfies
\be
k_{A'}^\mu {\cal M}_\mu = 0,
\ee
where ${\cal M}_\mu$ represents the Feynmann amplitude without the external
polarization vector.

The matrix elements for each diagram are given by
\begin{eqnarray}\nonumber
\mathcal{M}_a &&= i^2\frac{4m_N^2}{m_{\pi}^2}\frac{f_{pp}^2e\epsilon}{(k^2-m_{\pi}^2)}\frac{1}{(m_{A'}^2 +2(k_{A'}\cdot p_3))}\bar{u}(p_4)\gamma_5u(p_2)\bar{u}(p_3)\slashed{\epsilon}(\slashed{p}_3+\slashed{k}_{A'}+m_N)\gamma_5u(p_1)\\\nonumber
\mathcal{M}_a' &&= i^2\frac{4m_N^2}{m_{\pi}^2}\frac{f_{pp}^2e\epsilon}{(l^2-m_{\pi}^2)}\frac{1}{(m_{A'}^2 +2(k_{A'}\cdot p_4))}\bar{u}(p_3)\gamma_5u(p_2)\bar{u}(p_4)\slashed{\epsilon}(\slashed{p}_4+\slashed{k}_{A'}+m_N)\gamma_5u(p_1)\\\nonumber
\mathcal{M}_b &&= i^2\frac{4m_N^2}{m_{\pi}^2}\frac{f_{pp}^2e\epsilon}{(k^2-m_{\pi}^2)}\frac{1}{(m_{A'}^2 +2(k_{A'}\cdot p_4))}\bar{u}(p_3)\gamma_5u(p_1)\bar{u}(p_4)\slashed{\epsilon}(\slashed{p}_4+\slashed{k}_{A'}+m_N)\gamma_5u(p_2)\\\nonumber
\mathcal{M}_b' &&= i^2\frac{4m_N^2}{m_{\pi}^2}\frac{f_{pp}^2e\epsilon}{(l^2-m_{\pi}^2)}\frac{1}{(m_{A'}^2 +2(k_{A'}\cdot p_3))}\bar{u}(p_4)\gamma_5u(p_1)\bar{u}(p_3)\slashed{\epsilon}(\slashed{p}_3+\slashed{k}_{A'}+m_N)\gamma_5u(p_2)\\\nonumber
\mathcal{M}_c &&= i^2\frac{4m_N^2}{m_{\pi}^2}\frac{f_{pp}^2e\epsilon}{(k^2-m_{\pi}^2)}\frac{1}{(m_{A'}^2 -2(k_{A'}\cdot p_1))}\bar{u}(p_4)\gamma_5u(p_2)\bar{u}(p_3)\gamma_5(\slashed{p}_1-\slashed{k}_{A'}+m_N)\slashed{\epsilon}u(p_1)\\\nonumber
\mathcal{M}_c' &&= i^2\frac{4m_N^2}{m_{\pi}^2}\frac{f_{pp}^2e\epsilon}{(l^2-m_{\pi}^2)}\frac{1}{(m_{A'}^2 -2(k_{A'}\cdot p_1))}\bar{u}(p_3)\gamma_5u(p_2)\bar{u}(p_4)\gamma_5(\slashed{p}_1-\slashed{k}_{A'}+m_N)\slashed{\epsilon}u(p_1)\\\nonumber
\mathcal{M}_d &&= i^2\frac{4m_N^2}{m_{\pi}^2}\frac{f_{pp}^2e\epsilon}{(k^2-m_{\pi}^2)}\frac{1}{(m_{A'}^2 -2(k_{A'}\cdot p_2))}\bar{u}(p_3)\gamma_5u(p_1)\bar{u}(p_4)\gamma_5(\slashed{p}_2-\slashed{k}_{A'}+m_N)\slashed{\epsilon}u(p_2)\\\nonumber
\mathcal{M}_d' &&= i^2\frac{4m_N^2}{m_{\pi}^2}\frac{f_{pp}^2e\epsilon}{(l^2-m_{\pi}^2)}\frac{1}{(m_{A'}^2 -2(k_{A'}\cdot p_2))}\bar{u}(p_4)\gamma_5u(p_1)\bar{u}(p_3)\gamma_5(\slashed{p}_2-\slashed{k}_{A'}+m_N)\slashed{\epsilon}u(p_2)
\end{eqnarray}
where $f_{pp}$ is the coupling for the $p-p-\pi$ vertex, and we have used the momenta definitions $k \equiv p_2 -p_4$ and $l \equiv p_2 - p_3$.  

With these definitions we find the kinematic relations
\begin{eqnarray}\nonumber
p_1\cdot p_2 &=& -k\cdot l + m_N^2 -\frac{l^2}{2}-\frac{k^2}{2}+\frac{\ma^2}{2}\\\nonumber
p_1\cdot p_3 &=&m^2 + \frac{\ma^2}{2}-\frac{k^2}{2}\\\nonumber
p_1\cdot p_4 &=& m_N^2 + \frac{\ma^2}{2}-\frac{l^2}{2}\\\nonumber
p_2\cdot p_3 &=& m_N^2-\frac{l^2}{2}\\\nonumber
p_2\cdot p_4 &=& m_N^2-\frac{k^2}{2}\\\nonumber
p_3\cdot p_4 &=& k\cdot l + m_N^2 -\frac{l^2}{2}-\frac{k^2}{2}
\end{eqnarray}
As in previous work with massive axions \cite{Giannotti:2005tn}, we will work in the approximation that the nucleon mass is much larger than both the temperature and the vector mass, and that the direct and exchange four-momenta transferred is larger than the vector mass and momenta, $k^2, l^2 \gg \ka^2, \ma^2,\oa^2$.  This is due to the fact that the kinetic energy of the vector particles should roughly be a factor of a few times the temperature.  We can then use the simplifying relation $p_i\cdot \ka \simeq m_N\oa$.  Finally, we use the approximation $m_N^2 \gg |{\bf k}|^2$, since $|{\bf k}|^2 \simeq 3m_NT \ll m_N^2$.

Squaring the matrix elements, we find for the process $p + p \rightarrow p + p + A'$ 
\begin{eqnarray}
\label{eq:ampp2}
\sum_s|\feynm|^2_{p+p}=\frac{m_N^2}{m_{\pi}^4}\frac{64{\bf k}^2}{\oa^2}\left(\frac{C_{k}|{\bf k}|^4}{({\bf k}^2 + m_{\pi}^2)({\bf k}^2 + m_{\pi}^2)}+\frac{C_{l}|{\bf l}|^4}{({\bf l}^2 + m_{\pi}^2)({\bf l}^2 + m_{\pi}^2)}+\frac{C_{kl}(|{\bf k}|^2|{\bf l}|^2-2|{\bf k}\cdot{\bf l}|^2)}{({\bf k}^2 + m_{\pi}^2)({\bf l}^2 + m_{\pi}^2)}\right)
\end{eqnarray}
where $C_k = f_{pp}^4(g_{\alpha}^2+g_{\beta}^2)$, $C_l = f^4_{pp}(g_{\alpha}^2+g_{\beta}^2-2g_{\alpha}g_{\beta})$, and $C_{kl} = f_{pp}^4(g_{\alpha}^2+g_{\beta}^2-2g_{\alpha}g_{\beta})$.  Here the $g_i$ are the couplings between the $A'$ and the $i$'th particle; which in this case is $g_{\alpha}=g_{\beta} \equiv g_p = e\epsilon$.  Thus, $C_k$ will be the only non-zero contribution.

\section{Phase space integrations for the process 
$ p + p \rightarrow p + p + A'$}

The energy emission rate is given by
\begin{eqnarray}
\label{eq:emrate}
Q_{A} = \int d\Pi \ S \sum_{spins}|\mathcal{M}|^2(2\pi)^4\oa\delta^4(p_1+p_2-p_3-p_4-\ka)f_1f_2,
\end{eqnarray}
where
\begin{eqnarray}
d\Pi = \frac{d^3{\bf p}_1}{(2\pi)^32E_1}\frac{d^3{\bf p}_2}{(2\pi)^32E_2}\frac{d^3{\bf p}_3}{(2\pi)^32E_3}\frac{d^3{\bf p}_4}{(2\pi)^32E_4}\frac{d^3\bka}{(2\pi)^32\oa}.
\end{eqnarray}
The initial-state nucleon occupation numbers are given by the 
non-relativistic Maxwell-Boltzmann distribution 
\begin{eqnarray}
f({\bf p}) = \frac{n_B}{2}\left(\frac{2\pi}{\mn T}\right)^{3/2}
{\rm{e}}^{-{\bf p}^2/2\mn T},
\end{eqnarray}
and $S$ is the symmetry factor, $S = 1/4$ for the $pp$ process and $S = 1$ for the $np$ process.  In the non-relativistic approximation, $E_i \simeq \mn$.

Following~\cite{Brinkmann:1988vi,Raffelt:1993ix,Giannotti:2005tn} we go to 
the center-of-momentum coordinates
\begin{eqnarray}
{\bf p}_1 &\equiv& {\bf P} + {\bf p}_i\\
{\bf p}_2 &\equiv& {\bf P} - {\bf p}_i\\
{\bf p}_3 &\equiv& {\bf P}' + {\bf p}_f\\
{\bf p}_4 &\equiv& {\bf P} '- {\bf p}_f.
\end{eqnarray}
In the non-relativistic limit, a typical nucleon with kinetic energy
$E_{kin}$ has momentum $|{\bf p}|
\approx \sqrt{2 \mn E_{kin}} \gg E_{kin} \approx \bka$. 
Then, the three dimensional delta function will enforce ${\bf P} = {\bf P}'$. 

One defines the dimensionless variables
\begin{eqnarray}
u \equiv \frac{{\bf p}_i^2}{\mn T}\,\,\, v \equiv 
\frac{{\bf p}_f^2}{\mn T}\,\,\, 
x \equiv \frac{\oa}{T}\,\,\,
y \equiv \frac{m_{\pi}^2}{\mn T}\,\,\, 
q \equiv \frac{\ma}{T},
\end{eqnarray}
as well as $z\equiv \cos(\theta_{if})$, where $\theta_{if}$ is the angle between ${\bf p}_i$ and ${\bf p}_f$.
The velocity of the emitted boson, $\zeta$, can be written as
\begin{eqnarray}
\zeta^2 = \frac{{\bka}^2}{\oa^2} = (1-\frac{q^2}{x^2}).
\end{eqnarray}
The momenta in Eq.~(\ref{eq:ampp2}) can then be written as
\begin{eqnarray}
|{\bf k}|^2 &=& \mn T(u+v-2z\sqrt{uv})\\
|{\bf l}|^2 &=& \mn T(u+v+2z\sqrt{uv})\\
|{\bf k}|^2+m_{\pi}^2 &=& \mn T(u+v-2z\sqrt{uv} +y)\\
|{\bf l}|^2+m_{\pi}^2 &=& \mn T(u+v+2z\sqrt{uv} +y)\\
|{\bf k}\cdot {\bf l}|^2 &=& \left(\mn T\right)^2(u-v)^2.
\end{eqnarray}
The delta function in the emission rate, Eq.~(\ref{eq:emrate}), becomes
\begin{eqnarray}
\delta\left(p_1 + p_2 - p_3 -p_4 -\ka\right) 
= \delta^{(3)}\left({\bf p}_1+{\bf p}_2-{\bf p}_3-{\bf p}_4-\bka\right)
\delta(u-v-x)/T,
\end{eqnarray}
and we use the three-dimensional piece to do the ${\bf p}_4$ integration.

We now change coordinate bases from ${\bf p}_1$, ${\bf p}_2$, ${\bf p}_3$ 
to the ${\bf p}_i$, ${\bf p}_f$, ${\bf P}$ system, including a factor of 8 
(the inverse Jacobian).

Using
\begin{eqnarray}
f_1f_2 = \frac{n_B^2}{4}\left(\frac{2\pi}{\mn T}\right)^3{\rm{e}}^{-u}
{\rm{e}}^{-{\bf P}^2/\mn T},
\end{eqnarray}
we obtain
\begin{eqnarray}
\int d^3{\bf p}_i{\rm{e}}^{-{\bf p}_i^2/\mn T} &=& 4\pi \int {\bf p}_i^2 
{\rm{e}}^{-{\bf p}_i^2/\mn T} d{\bf p}_i = 
2\pi (\mn T)^{3/2}\int \sqrt{u}{\rm{e}}^{-u}du\\
\int d^3{\bf p}_f &=& 2\pi \int {\bf p}_f^2 d{\bf p}_f\int_{-1}^{1}dz = \pi 
(\mn T)^{3/2}\int \sqrt{v}dv\int_{-1}^1dz\\
\int d^3{\bf P}{\rm{e}}^{-{\bf P}^2/\mn T} &=& 4\pi\int P^2{\rm{e}}^{-P^2/\mn T}dP = 
4\pi (\mn T)^{3/2}\frac{\sqrt{\pi}}{4}.
\end{eqnarray}

Gathering the factors of $\oa$ in the initial emission expression,
Eq.~(\ref{eq:emrate}), the denominator from the phase space
integration, and the amplitude squared, Eq.~(\ref{eq:ampp2}), we are left
with
\begin{eqnarray}
\int d^3{\bka }\frac{1}{\oa^2}.
\end{eqnarray}
Using
\begin{eqnarray}
{\bka}^2 = \zeta^2\oa^2
\end{eqnarray}
we have
\begin{eqnarray}
\int d^3{\bka}\frac{1}{\oa^2}=4\pi\int d{\bka} \frac{{\bka}^2}{\oa^2} 
=4\pi\int d{\bka}\zeta^2 = 4\pi T\int dx\frac{\sqrt{x^2-q^2}}{x} 
\end{eqnarray}

Using the above relations for the phase space integration factors, we insert $|\mathcal{M}|^2$ given in Eq.(\ref{eq:ampp2}) into Eq.(\ref{eq:emrate}), and find that the emission rate for the $p + p \rightarrow p + p + A'$ process is given by
\begin{eqnarray}\label{qpp}
Q_{pp} = \frac{\alpha_{\pi}^2T^{2.5}n_B^2}{32\pi^{1.5}m_N^{3.5}}\int du dv dz dx \sqrt{uv}{\rm{e}}^{-u}\frac{\sqrt{x^2-q^2}}{x}\delta(u-v-x)\times(I_k+I_l +I_{kl}+I_{k\cdot l})
\end{eqnarray}
where $\alpha_{\pi} \equiv (2m_N/f_{pp})^2/4\pi$ and the various integrand pieces are defined as
\begin{eqnarray}
I_k &=& f^4(g_{\alpha}^2+g_{\beta}^2)\frac{(u+v-2z\sqrt{uv})^3}{(u+v-2z\sqrt{uv}+y)^2}\\\nonumber
I_l &=& f^4(g_{\alpha}^2+g_{\beta}^2-2g_{\alpha}g_{\beta})\frac{(u+v+2z\sqrt{uv})^2}{(u+v+2z\sqrt{uv}+y)^2}(u+v-2z\sqrt{uv})\\\nonumber
I_{kl} &=& f^4(g_{\alpha}^2+g_{\beta}^2-2g_{\alpha}g_{\beta})\frac{(u+v)^2-4z^2uv)}{(u+v+y)^2-4z^2uv}(u+v-2z\sqrt{uv})\\\nonumber
I_{k\cdot l} &=& -2f^4(g_{\alpha}^2+g_{\beta}^2-2g_{\alpha}g_{\beta})\frac{(u-v)^2}{(u+v+y)^2-4z^2uv}(u+v-2z\sqrt{uv})
\end{eqnarray}

\section{Amplitude for the process $p + n\rightarrow p+ n + A'$}

The requisite matrix elements for the process $p + n\rightarrow p + n + A'$ are 
\begin{eqnarray}
\mathcal{M}_a &=&\frac{1}{|{\bf k}|^2 + m_{\pi}^2}\frac{1}{2m_N\oa}\bigg(\frac{2m_N}{m_{\pi}}\bigg)^2g_{p}f_{pn}^2\bar{u}_3\slashed{\epsilon}(\slashed{p}_3+\slashed{k}_{A'}+m_N)\gamma^5u_1\bar{u}_4\gamma^5u_2\\
\mathcal{M}_c&=&-\frac{1}{|{\bf k}|^2 + m_{\pi}^2}\frac{1}{2m_N\oa}\bigg(\frac{2m_N}{m_{\pi}}\bigg)^2g_{p}f_{pn}^2\bar{u}_3\gamma^5(\slashed{p}_1-\slashed{k}_{A'}+m_N)\slashed{\epsilon}u_1\bar{u}_4\gamma^5u_2\\
\mathcal{M}_{b'}&=&\frac{1}{|{\bf l}|^2 + m_{\pi}^2}\frac{1}{2m_N\oa}\bigg(\frac{2m_N}{m_{\pi}}\bigg)^2g_{p}f_{pn}^2\bar{u}_3\slashed{\epsilon}(\slashed{p}_3+\slashed{k}_{A'}+m_N)\gamma^5u_2\bar{u}_4\gamma^5u_1\\
\mathcal{M}_{c'}&=&-\frac{1}{|{\bf l}|^2 + m_{\pi}^2}\frac{1}{2m_N\oa}\bigg(\frac{2m_N}{m_{\pi}}\bigg)g_{p}f_{pn}^2\bar{u}_4\gamma^5(\slashed{p}_1-\slashed{k}_{A'}+m_N)\slashed{\epsilon}u_1\bar{u}_3\gamma^5u_2\\
\mathcal{M}_{e'}&=& \frac{1}{|{\bf l}|^2 + m_{\pi}^2}\frac{1}{(l-a)^2-m_{\pi}^2}\bigg(\frac{2m_N}{m_{\pi}}\bigg)^2g_{\pi}f_{pn}^2\bar{u}_4\gamma^5u_1\bar{u}_3\gamma^5u_2(k_{A'}-2l)\cdot \epsilon
\end{eqnarray}
where the coupling $f_{pn}$ is the coupling for the $p-n-\pi$ vertex, which is related to $f_{pp}$ as $f_{pn} = \sqrt{2}f_{pp}$ by isospin invariance.  As in the $pp$ case, $g_{i}$ is the coupling between the $A'$ and the $i$'th particle.  For this process we have $g_{\pi} = g_p = e\epsilon$.

The products of matrix elements for diagrams with bremsstrahlung originating off of external legs will produce terms in $|\mathcal{M}|^2$ identical to those found in Appendix A for the $p + p \rightarrow p + p + A'$ process.  However, we see that we will also find matrix element products which include the factor $\mathcal{M}_{E'}$ which arises from internal bremsstrahlung.  These new combinations are given by
\begin{eqnarray}\nonumber
\mathcal{M}_{e'}\mathcal{M}_{e'}^{\dag}&=&-\alpha_{e'e'}\textrm{Tr}[(\slashed{p}_4+m_N)\gamma^5(\slashed{p}_1+m_N)\gamma^5]\textrm{Tr}[(\slashed{p}_2+m_N)\gamma^5(\slashed{p}_3+m_N)\gamma^5]\\&\times&(k_{A'}-2l)\cdot (k_{A'}-2l)\\
\alpha_{e'e'} &=& \frac{1}{(|{\bf l}|^2+m_{\pi}^2)^2}\frac{1}{((l-k_{A'})^2-m_{\pi}^2)^2}\frac{16m_N^4}{m_{\pi}^4}g_{\pi}^2f_{pn}^4\\
\mathcal{M}_{a}\mathcal{M}_{e'}^{\dag}&=&-\alpha_{ae'}\\\nonumber&\times&Tr[(\slashed{p}_3+m_N)(\slashed{k}_{A'}-2\slashed{l})(\slashed{p}_3+\slashed{k}_{A'}+m_N)\gamma^5(\slashed{p}_1+m_N)\gamma^5(\slashed{p}_4+m_N)\gamma^5(\slashed{p}_2+m_N)\gamma^5]\\
\alpha_{ae'} &=& \frac{1}{(|{\bf l}|^2+m_{\pi}^2)}\frac{1}{(|{\bf k}|^2+m_{\pi}^2)}\frac{1}{((l-k_{A'})^2-m_{\pi}^2)}\frac{8m_N^3}{\oa m_{\pi}^4}g_{\pi}g_pf_{pn}^4\\
\mathcal{M}_{c}\mathcal{M}_{e'}^{\dag}&=&-\alpha_{ce'}\\\nonumber&\times&Tr[(\slashed{p}_3+m_N)\gamma^5(\slashed{p}_1-\slashed{k}_{A'}+m_N)(\slashed{k}_{A'}-2\slashed{l})(\slashed{p}_1+m_N)\gamma^5(\slashed{p}_4+m_N)\gamma^5(\slashed{p}_2+m_N)\gamma^5]\\\\
\alpha_{ce'} &=& -\frac{1}{(|{\bf l}|^2+m_{\pi}^2)}\frac{1}{(|{\bf k}|^2+m_{\pi}^2)}\frac{1}{((l-k_{A'})^2-m_{\pi}^2)}\frac{8m_N^3}{\oa m_{\pi}^4}g_{\pi}g_pf_{pn}^4\\
\mathcal{M}_{b'}\mathcal{M}_{e'}^{\dag}&=&-\alpha_{b'e'}\textrm{Tr}[(\slashed{p}_3+m_N)(\slashed{k}_{A'}-2\slashed{l})(\slashed{p}_3+\slashed{k}_{A'}+m_N)\gamma^5(\slashed{p}_2+m_N)\gamma^5]\\\nonumber&\times&\textrm{Tr}[(\slashed{p}_4+m_N)\gamma^5(\slashed{p}_1+m_N)\gamma^5)]\\
\alpha_{b'e'} &=& \frac{1}{(|{\bf l}|^2+m_{\pi}^2)^2}\frac{1}{((l-k_{A'})^2-m_{\pi}^2)}\frac{8m_N^3}{\oa m_{\pi}^4}g_{\pi}g_pf_{pn}^4\\
\mathcal{M}_{c'}\mathcal{M}_{e'}^{\dag}&=&-\alpha_{c'e'}\textrm{Tr}[(\slashed{p}_4+m_N)\gamma^5(\slashed{p}_1-\slashed{k}_{A'}+m_N)(\slashed{k}_{A'}-2\slashed{l})(\slashed{p}_1+m_N)\gamma^5]\\\nonumber&\times&\textrm{Tr}[(\slashed{p}_3+m_N)\gamma^5(\slashed{p}_2+m_N)\gamma^5)]\\
\alpha_{c'e'} &=& -\frac{1}{(|{\bf l}|^2+m_{\pi}^2)^2}\frac{1}{((l-k_{A'})^2-m_{\pi}^2)}\frac{8m_N^3}{\oa m_{\pi}^4}g_{\pi}g_pf_{pn}^4\\
\end{eqnarray}

\begin{figure}
\begin{center}
\includegraphics[width=0.7\textwidth]{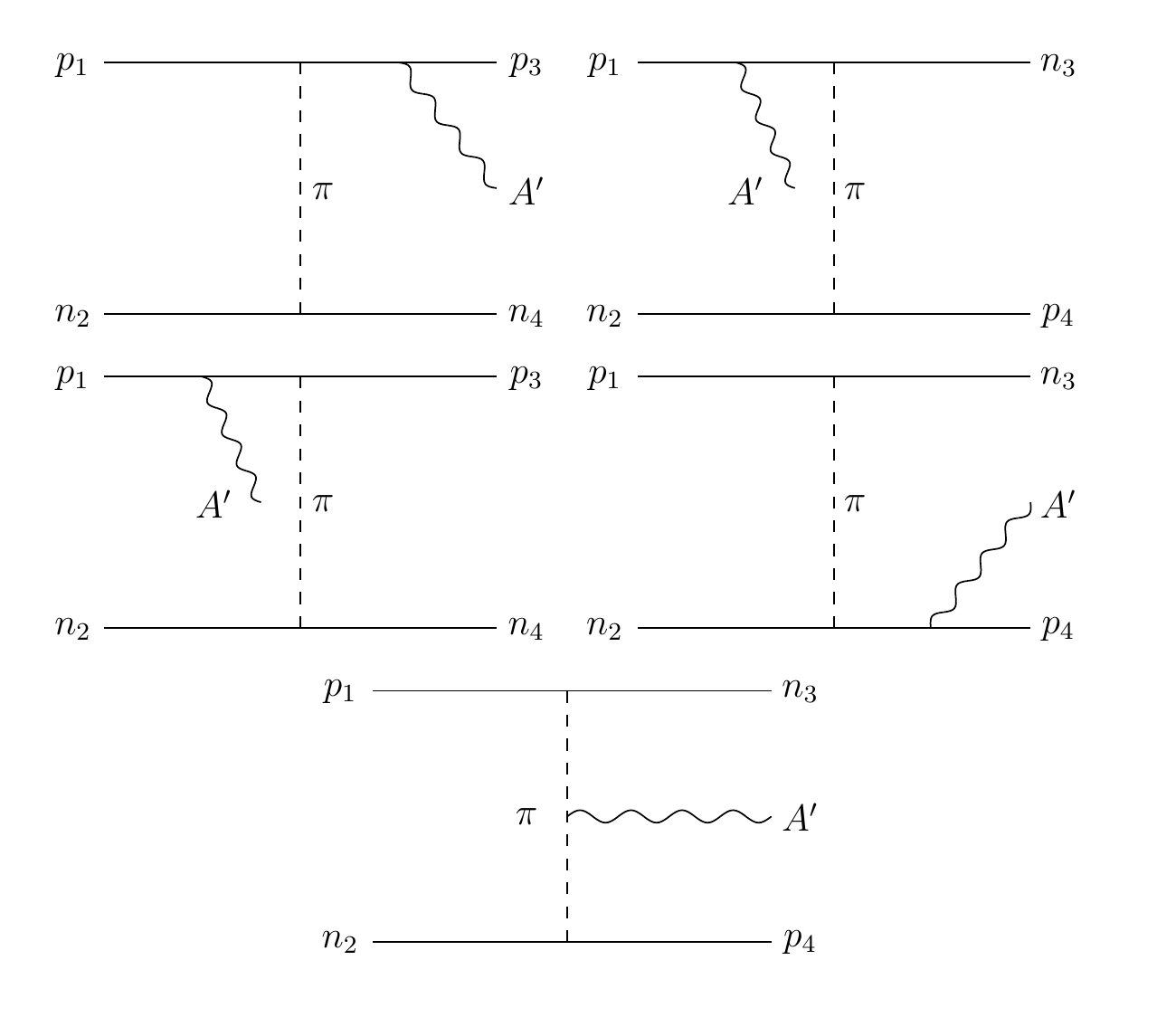}
\end{center}
\caption{Diagrams for the process $p + n \rightarrow p + n + A'$.  From top left to bottom left, the external bremsstrahlung diagrams are labelled $a$ and $c$, while from top right to bottom right, the external leg bremsstrahlung diagrams are labelled $b'$ and $c'$.  The internal bremsstrahlung process is labelled $e'$.}
\label{fig:feynman2}
\end{figure}

Putting everything together we find
\begin{eqnarray}\nonumber
\sum_s \mathcal{M}\mathcal{M}^{\dag}_{np\rightarrow npa} &=& \frac{64m_N^2|{\bf k}|^2}{\oa^2m_{\pi}^4}\bigg(\frac{C_k|{\bf k}|^4}{(|{\bf k}|^2 + m_{\pi}^2)^2}+\frac{C_l|{\bf l}|^4}{(|{\bf l}|^2 + m_{\pi}^2)^2}\bigg)\\\nonumber
&+& \frac{128m_N^3|{\bf k}\cdot {\bf l}|}{\oa m_{\pi}^4}\bigg(\frac{C_{AE'}(|{\bf k}|^2|{\bf l}|^2-2|{\bf k}\cdot {\bf l}|^2)}{(|{\bf k}|^2 + m_{\pi}^2)(|{\bf l}|^2 + m_{\pi}^2)((l-k_{A'})^2-m_{\pi}^2)}\\\nonumber
&+&\frac{2C_{B'E'}|{\bf l}|^4}{(|{\bf l}|^2 + m_{\pi}^2)^2((l-k_{A'})^2-m_{\pi}^2)}\bigg)\\
&+&\frac{64m_N^4}{m_{\pi}^4}\frac{C_{E'E'}|{\bf l}|^6}{(|{\bf l}|^2 + m_{\pi}^2)^2((l-k_{A'})^2-m_{\pi}^2)^2}
\end{eqnarray}
where
\begin{eqnarray}
C_k &\equiv& f_{pn}^4g_p^2=f_{pn}^4e^2\epsilon^2\\
C_l &\equiv& f_{pn}^4g_p^2=f_{pn}^4e^2\epsilon^2\\
C_{AE'} &\equiv& g_{\pi}g_pf_{pn}^4=f_{pn}^4e^2\epsilon^2\\
C_{B'E'} &\equiv& g_{\pi}g_pf_{pn}^4=f_{pn}^4e^2\epsilon^2\\
C_{E'E'} &\equiv &g_{\pi}^2f_{pn}^4=f_{pn}^4e^2\epsilon^2
\end{eqnarray}

\section{Phase Space Integrations for $n + p \rightarrow n + p + A'$}

We can now compute the phase space integrations in the same manner as in the $p + p \rightarrow p + p + A'$ case.  We find
\begin{eqnarray}\label{qpn}
Q_{pn} = &&\frac{\alpha_{\pi}^2T^{2.5}n_B^2}{8\pi^{1.5}m_N^{3.5}}\int du dv dz dx\sqrt{uv}{\rm{e}}^{-u}\sqrt{1-\frac{q^2}{x^2}}\delta(u-v-x)\\&\times&(I_k+I_l +I_{AE'kl}+I_{AE'k\cdot l}+I_{B'E'}+I_{E'E'})
\end{eqnarray}
where (note the $x$ dependence of the last four)
\begin{eqnarray}
I_k &=& C_k\frac{(u+v-2z\sqrt{uv})^3}{(u+v-2z\sqrt{uv}+y)^2}\\\nonumber
I_l &=& C_l\frac{(u+v+2z\sqrt{uv})^2}{(u+v+2z\sqrt{uv}+y)^2}(u+v-2z\sqrt{uv})\\\nonumber
I_{AE'kl} &=& 2xC_{AE'}\frac{((u+v)^2-4z^2uv)}{((u+v+y)^2-4z^2uv)}\frac{(u-v)}{(u+v+y+2z\sqrt{uv})}\\\nonumber
I_{AE'k\cdot l} &=& -4xC_{AE'}\frac{(u-v)^3}{((u+v+y)^2-4z^2uv)}\frac{1}{(u+v+y+2z\sqrt{uv})}\\\nonumber
I_{B'E'} &=& 4xC_{B'E'}\frac{(u+v+2z\sqrt{uv})^2}{(u+v+2z\sqrt{uv}+y)^3}(u-v)\\\nonumber
I_{E'E'}&=& x^2C_{E'E'}\frac{(u+v+2z\sqrt{uv})^3}{(u+v+2z\sqrt{uv}+y)^4}
\end{eqnarray}

\bibliography{darkbosonbrem}
\bibliographystyle{h-physrev5}

\end{document}